\documentclass[journal]{IEEEtran}
\usepackage{subcaption}
\usepackage{cite}
\usepackage{tabularray}
\usepackage{graphicx}
\usepackage{soul}
\usepackage{amsmath,amssymb,amsfonts}
\usepackage{algorithmic}
\usepackage{textcomp}
\usepackage{placeins}
\usepackage{float}
\usepackage{dblfloatfix}
\usepackage{array}
\usepackage{enumitem}
\usepackage{lscape}
\usepackage{multirow}
\usepackage[table,xcdraw]{xcolor}
\usepackage{url}
\usepackage{changepage}
\usepackage{makecell}
\usepackage{caption}
\usepackage{longtable}
\usepackage[normalem]{ulem}

\begin{document}

\title{Towards Resilient Intrusion Detection in CubeSats: \\ Challenges, TinyML Solutions, and Future Directions}

\author{Yasamin Fayyaz, Li Yang, \IEEEmembership{Member, IEEE}, and Khalil El-Khatib, \IEEEmembership{Member, IEEE}%
\thanks{Manuscript published in IEEE Aerospace and Electronic Systems Magazine in March 2026.}%
\thanks{All authors are with the Faculty of Business and IT, Ontario Tech University, Oshawa, Ontario, L1G 0C5, Canada.%
\newline Corresponding author: Yasamin Fayyaz (yasamin.fayyaz@ontariotechu.net). \newline Other authors' contact information: Li Yang (li.yang@ontariotechu.ca), and Khalil El-Khatib (khalil.el-khatib@ontariotechu.ca).}%
}

\maketitle

\begin{abstract} 
CubeSats have revolutionized access to space by providing affordable and accessible platforms for research and education. However, their reliance on Commercial Off-The-Shelf (COTS) components and open-source software has introduced significant cybersecurity vulnerabilities. Ensuring the cybersecurity of CubeSats is vital as they play increasingly important roles in space missions. Traditional security measures, such as intrusion detection systems (IDS), are impractical for CubeSats due to resource constraints and unique operational environments. This paper provides an in-depth review of current cybersecurity practices for CubeSats, highlighting limitations and identifying gaps in existing methods. Additionally, it explores non-cyber anomaly detection techniques that offer insights into adaptable algorithms and deployment strategies suitable for CubeSat constraints. Open research problems are identified, including the need for resource-efficient intrusion detection mechanisms, evaluation of IDS solutions under realistic mission scenarios, development of autonomous response systems, and creation of cybersecurity frameworks. The addition of TinyML into CubeSat systems is explored as a promising solution to address these challenges, offering resource-efficient, real-time intrusion detection capabilities. Future research directions are proposed, such as integrating cybersecurity with health monitoring systems, and fostering collaboration between cybersecurity researchers and space domain experts.
\end{abstract}


\section{Introduction}
\begin{figure}[h]
    \centering
    \includegraphics[width=1\columnwidth]{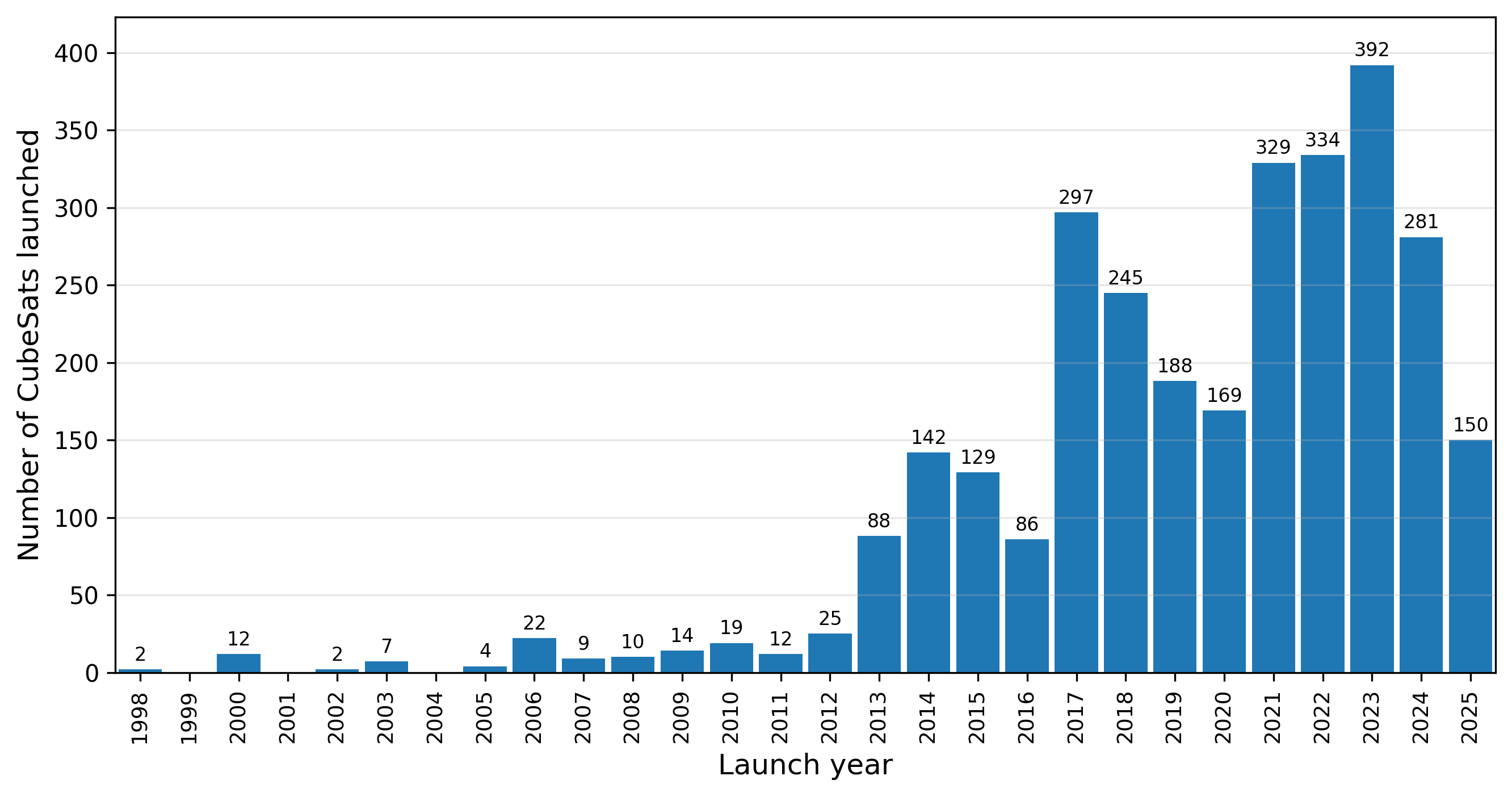}
    \caption{Annual CubeSat launches from 1998 to 2025. Counts were compiled from the Nanosatellite and CubeSat Database \cite{eusat}.}
    \label{fig:launches}
\end{figure}

CubeSats are a type of nanosatellite that weigh less than 1.33 kg per unit and have dimensions of 10 cm x 10 cm x 10 cm \cite{nasa_cubesat_101}. These small, cost-effective satellites have made space research and education more accessible by providing affordable access to space. As shown in Fig. \ref{fig:launches}, annual CubeSat launches were typically in the single digits across the late 1990s and early 2000s, then increased sharply and stabilized at hundreds of launches per year in subsequent years \cite{eusat}. This rise in popularity is primarily due to the availability of Commercial Off-The-Shelf (COTS) components, lower costs, and the evolution of launch services, which have become more frequent and flexible, allowing for more opportunities to send CubeSats into space \cite{physics_based}. For example, NASA's CubeSat Launch Initiative (CSLI) supports the launch of CubeSats by providing opportunities to fly as auxiliary payloads on future launches with excess capacity or as deployments from the International Space Station (ISS) \cite{nasa_cubesat_101}. Additionally, ground station as a service (GSaaS) platforms, like those provided by AWS \cite{aws_ground_station}, further reduce operational costs and complexity. CubeSats play an important role in various missions, including environmental monitoring, disaster prevention, communications, scientific research, and military applications \cite{applications, ids_survey}, highlighting their importance in advancing space technology and expanding our understanding of space.

The growing number of CubeSats raises significant cybersecurity concerns. The ground segment, link segment, and space segment are all subject to cyberattacks, which can range from data tampering to malicious radio commands that disable the CubeSat \cite{attack_tree}. Traditionally, security through obscurity, where protection is assumed because on-board firmware and interfaces are kept inaccessible to hinder external analysis and exploitation, was a common approach in space systems \cite{willbold2023space}.
However, this strategy is ineffective for CubeSats due to the widespread use of Commercial Off-The-Shelf (COTS) components and open-source software. While these widely available technologies reduce development costs and lower barriers to entry, they also increase exposure, as attackers can readily study and exploit them \cite{physics_based}.
Consequently, the need for robust and efficient cybersecurity measures for CubeSats has become increasingly critical.

In terrestrial settings, intrusion detection systems (IDS), which monitor system activities for signs of cyber anomalies, have long been used as a cybersecurity defense tool to identify and mitigate potential threats. These systems analyze various data sources, such as log files and network traffic to detect intrusions and ensure the integrity and security of information systems \cite{ids_survey}. For CubeSats, an IDS can be implemented using two main architectures: host-based and network-based. A host-based IDS focuses on monitoring activity on an individual endpoint rather than monitoring traffic across the full network \cite{computers14030087}. It would analyze system activities and log files within the CubeSat itself, offering autonomous protection and monitoring. On the other hand, a network-based IDS monitors traffic traversing network links by analyzing packets and flows \cite{computers14030087}; in CubeSat missions this monitoring would be performed at ground-station gateways where traffic from multiple nodes converges, providing visibility across multi-satellite systems \cite{ids_survey}.
However, traditional IDSs are impractical for satellite systems due to resource constraints and their operational environment. In general, space systems often lack the computational and power resources needed to support traditional IDSs. This is especially true for CubeSats, which are even more resource-constrained than larger satellites, including limitations in computational and power resources \cite{ids_survey}. Additionally, the metrics and methods used for anomaly detection in terrestrial IT systems do not work well in space. Establishing a baseline behavior for satellites before deployment is difficult due to the many variables and changes in the space environment. Furthermore, the baseline behavior of a satellite evolves over its mission life, complicating the use of traditional IDS methods \cite{physics_based}.

Machine learning (ML) models are increasingly used to develop IDSs across various domains due to their ability to detect both known and unknown attack patterns through anomaly detection \cite{ids_survey}. However, applying ML models used in terrestrial IDSs to CubeSats would introduce challenges such as high resource consumption and computational overhead \cite{ids_survey}. Given these limitations, effective intrusion detection in CubeSats requires innovative solutions. The resource constraints and constantly changing operational environments of CubeSats call for the development of lightweight, adaptive, and efficient detection mechanisms designed specifically for these systems. Traditional cybersecurity methods often fall short; thus, new approaches that consider the unique constraints of CubeSats are essential to defending these critical assets from evolving cyber threats.

This paper represents the first study to investigate the application of TinyML methods specifically for CubeSat cybersecurity, addressing a research gap and providing a structured foundation that can guide future efforts toward deploying TinyML-based security mechanisms in operational CubeSat systems. TinyML refers to the application of ML algorithms on small, resource-constrained devices like microcontrollers \cite{tinyML}. It is advantageous for CubeSats due to its low power consumption, small memory footprint, and ability to run efficiently on limited hardware \cite{tinyML}. This enables real-time data processing and host-based anomaly detection performed directly on the CubeSat.

This paper makes the following contributions to the field of CubeSat Cybersecurity:
\begin{itemize}
    \item Providing a comprehensive review of current cybersecurity practices specific to CubeSats.
    \item Exploring non-cyber anomaly detection methods, providing insights into adaptable algorithms and deployment techniques that can inform intrusion detection solutions under CubeSat-specific constraints.
    \item Investigating open research problems and synthesizing key limitations and gaps in current CubeSat cybersecurity and IDS research, with emphasis on challenges introduced by CubeSat-specific resource constraints.
    \item Introducing and examining the potential of TinyML for CubeSat cybersecurity, detailing its benefits, challenges, and practical applications in resource-constrained environments.
    \item Proposing potential future directions for research and development in CubeSat cybersecurity, including integrating cybersecurity with health monitoring systems and fostering collaboration between cybersecurity researchers and space experts.
    \item Establishing a foundational roadmap for applying TinyML to next-generation, on-board intrusion detection in satellites in general, using CubeSats as a common-denominator platform that captures widely shared resource constraints and operational realities \cite{attack_tree}.
\end{itemize}

The remainder of this paper is as follows: Section~\ref{sec:cyber} provides an overview of efforts towards the cybersecurity of CubeSats, explaining various vulnerabilities and existing measures. Section~\ref{sec:noncyber} explores non-cyber anomaly detection methods in CubeSats, discussing ML approaches that can be adapted for cybersecurity purposes. In Section~\ref{sec:openresearch}, open research problems in implementing effective IDSs within the strict constraints of CubeSat platforms are identified. Section~\ref{sec:tinyml} introduces the potential of TinyML for CubeSat cybersecurity, discussing its benefits, challenges, and practical applications in resource-constrained environments. Section~\ref{sec:future} outlines other future directions for research and development in CubeSat cybersecurity, focusing on approaches beyond TinyML. Finally, the conclusion summarizes the key findings and emphasizes the importance of advancing intrusion detection tailored to the challenges of CubeSats.

\section{Efforts Towards Cybersecurity of CubeSats}
\label{sec:cyber}
CubeSats, once considered a novelty in space missions, now play significant roles ranging from educational projects to national defense activities. Their affordability and rapid development cycles have popularized their use but have also introduced significant security challenges. The same factors that make CubeSats cost-effective (i.e., the use of Commercial Off-The-Shelf (COTS) components and open-source technologies) also make them vulnerable to cyberattacks \cite{attack_tree, hosted_payload, HACK}. This section explores the efforts aimed at enhancing CubeSat cybersecurity, including vulnerability assessments, threat simulations, advanced testing methodologies, and the development of real-time protective measures.

\subsection{Vulnerability Assessment and Attack Simulations}
The authors in \cite{attack_tree} used attack tree analysis to explore various vulnerabilities within CubeSat systems, including denial of service (DoS) attacks that prevent normal operations by disrupting communications or blocking required command execution, data tampering attacks that maliciously modify sensitive telemetry or sensor records stored in the ground station flight database, and attacks that disable the CubeSat by inserting and executing a malicious ``kill radio" command, providing a structured method to assess and mitigate potential cyber threats.

Expanding on specific cyber threats, the study in \cite{hosted_payload} investigated a simulated cyberattack targeting the primary payload of the OPS-SAT spacecraft, a CubeSat operated by the European Space \cite{opssat}. The attack targeted the CubeSat's primary payload by manipulating the camera system's executable. It delivered attacker-selected images whenever the program ran, via library hijacking, which tricks the executable into loading a malicious look-alike software library instead of the legitimate one. This scenario showed an example of the vulnerabilities present in satellite missions that incorporate Commercial Off-The-Shelf (COTS) components and rely on increasingly complex software systems.

Both studies contributed valuable insights into the cybersecurity challenges facing CubeSat operations. Falco \emph{et. al} \cite{attack_tree} were among the first to address CubeSat cybersecurity risks, which differ from traditional satellites due to increased democratization and accessibility, necessitating tailored security measures. While the paper effectively identified these vulnerabilities, a more comprehensive analysis of the practical challenges and limitations associated with implementing the proposed countermeasures in real-world CubeSat missions would strengthen its contribution. In contrast, the study in \cite{hosted_payload} stood out by moving beyond theoretical insights in CubeSat cybersecurity, successfully demonstrating a cyberattack targeting OPS-SAT and proving the vulnerability of CubeSats to low-resource attacks. However, certain aspects of the attack scenario remained unexplored, such as whether the malicious library would be generated on the satellite or on the ground. This uncertainty may limit both reproducibility and real-world applicability.

\subsection{Simulation Platforms for Cybersecurity Testing}
To test and develop mitigation strategies against CubeSat vulnerabilities, the authors in \cite{HACK} proposed the use of high-fidelity digital models, or Digital Twins (DTs). The Digital Twin approach would enable the simulation of several "what-if" scenarios involving potential cyberattacks The Digital Twin approach enabled the simulation of several ``what-if'' scenarios involving potential cyberattacks like GNSS spoofing, where an attacker transmits counterfeit navigation signals so the receiver reports false position or time \cite{attack_tree}; signal interception, where an adversary captures satellite-link information such as an IP address to initiate an unauthorized connection \cite{HACK}; and satellite-to-satellite attacks, where one satellite directly targets another in orbit, for example using an electromagnetic pulse \cite{HACK}. By creating a virtual prototype, the authors studied how cyber intrusions could affect CubeSat subsystems, potentially leading to mission failure due to cascading effects. Using a model-based systems engineering (MBSE) approach, the research emphasized the importance of subsystem interconnections and employed multiphysics simulations to analyze these interactions. Another initiative that has been developed to test and understand the cybersecurity vulnerabilities of CubeSats was the LinkStar Cybersecurity Sandbox, as introduced in \cite{linkstar}. The sandbox environment allowed for the simulation of cyberattacks across all segments of a satellite mission (ground, link, and space segments). Providing a controlled environment where cybersecurity professionals can test CubeSat components under simulated attack conditions.

The contributions of both papers advanced CubeSat cybersecurity. The study in \cite{HACK} provided insight into cybersecurity testing for CubeSats through an idealized digital twin environment. However, it lacked practical guidance on how to develop such a model, which made the feasibility of implementing a Digital Twin for CubeSat cybersecurity uncertain. The absence of implementation strategies limited the immediate applicability of this largely conceptual framework. In contrast, the study in \cite{linkstar} was valuable for promoting hands-on security testing, as cybersecurity researchers often lack practical testing platforms in the space domain. However, unlike the proposed digital twin model, which simulated cyberattacks within a multiphysics environment by accounting for various physical phenomena in the space system, this paper lacked detailed information on how it simulated environmental data. This is important due to CubeSats' constantly changing conditions in space, making replication of these factors to a reasonable degree essential for developing effective security solutions.

\subsection{Intrusion Detection Systems}
While testing and modeling platforms offered crucial insights and preventive measures, there remained a need for real-time detection and response systems. IDSs were designed to fulfill this role, providing an active layer of defense against evolving cyber threats. The authors in \cite{ids_survey} described the essential characteristics that an IDS for CubeSats should possess. These included access to multiple data sources, an adaptable system architecture, and various detection mechanisms. The IDS must balance security performance with the operational constraints of CubeSats, such as limited computational power and energy resources. The physics-informed intrusion detection system discussed in \cite{physics_based} proposed integrating the cyber state with the physical state of a CubeSat to provide more comprehensive detection capabilities. This approach would better address the challenges of establishing and adapting to changes in a satellite's baseline behavior throughout its mission life. The authors in \cite{scenario_based} explored IDSs using penetration testing on a notional CubeSat to generate attack scenarios that can disrupt satellite operations. The study considered various scenarios, such as memory and CPU consumption, imaging payload file deletion, log deletion, payload file deletion, and file partition removal, to exemplify how CubeSat's operations can be compromised.

The study in \cite{current_ids} introduced the first operational Intrusion Detection System specifically designed for CubeSats. This system was composed of a distributed architecture that used Artificial Neural Networks (ANNs) to detect malicious activities within the CubeSat Space Protocol (CSP) over the Controller Area Network (CAN). The IDS separated the detection into two stages: a time-based ANN classifier for analyzing packet timestamps and CAN IDs, and a data-based ANN classifier for inspecting packet payloads. The time-based model performed well in detecting DoS attacks and fuzzy injections, where an attacker introduces subtle variations in message payload or timing to disrupt CAN communication \cite{current_ids}, achieving F1-scores of 99.59\% and 90.23\%, respectively, while the data-based model identified replay attacks, where an adversary records legitimate CAN messages and later retransmits them to trigger unintended actions \cite{current_ids}, with an F1-score of 87.66\%.
To accommodate the resource limitations of CubeSats, training was performed on the ground segment, with periodic firmware updates to enhance on-board detection. Although no updated public information is available to confirm on-orbit deployment of the IDS, validated on an engineering model of a 3U CubeSat, the IDS was scheduled for space deployment in the second half of 2024 \cite{current_ids}.

The discussed studies contributed to the development of IDSs for CubeSats, though each had limitations. The survey by \cite{ids_survey} provided a realistic foundation for CubeSat IDS research by drawing on strategies from similar domains, such as drones and vehicles. However, these analogies might not perfectly address the unique environment of space. The physics-based approach in \cite{physics_based}, while innovative in integrating physical state data, lacked specific guidance on the practical combination of these data sources within CubeSat constraints. The scenario-based study in \cite{scenario_based} presented realistic attack scenarios that are valuable for testing IDS solutions. However, its focus on hypothetical situations limited its practical application to actual CubeSat architectures, which may vary significantly.

The operational IDS presented in \cite{current_ids}, while a significant step forward for CubeSat-specific cybersecurity, used the UAVCAN Attack Dataset \cite{uavcan} for training, collected from unmanned aerial vehicles (UAVs). Since UAVs and CubeSats operate in distinct environments with different communication patterns and system behaviors, training the IDS on UAV-derived data may not fully capture the unique characteristics of CubeSat operations. This reliance on a UAV-centric dataset raises concerns about the IDS's effectiveness when applied to CubeSat contexts, potentially resulting in reduced detection accuracy or higher false positives. This highlights the need for continuous refinement and testing of these systems to ensure effectiveness and practicality in space applications.

\begin{figure}[ht]
    \centering
    \includegraphics[width=1\columnwidth]{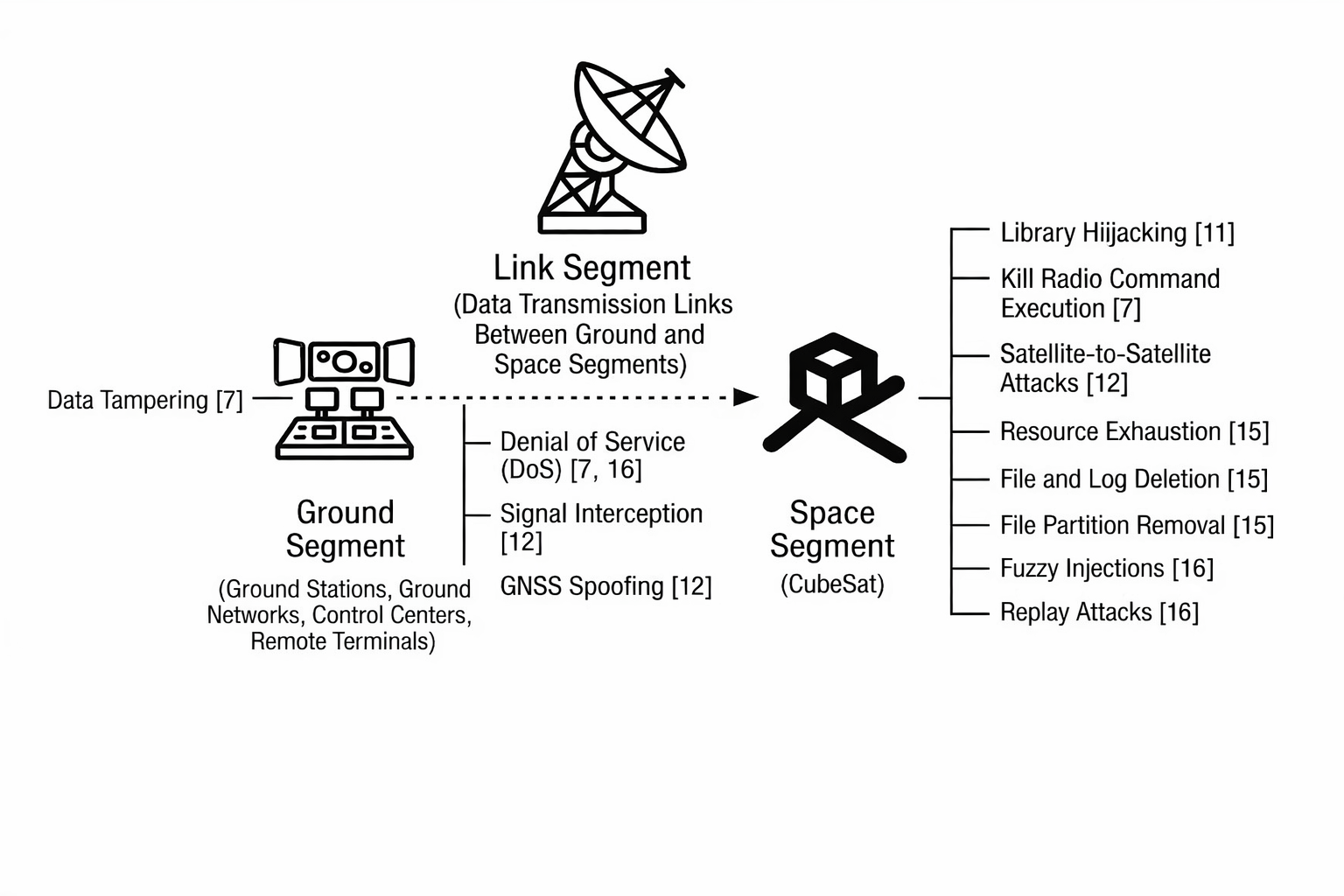}
    \caption{Taxonomy of cybersecurity threats and attacks targeting CubeSat mission segments, as reviewed in this section.}
    \label{fig:attacks_taxonomy}
\end{figure}

A consolidated overview of the studies reviewed in this section is provided in Table~\ref{tab:sec2-summary}. The table categorizes each work by approach, summarizes its purpose and primary technique, and highlights key contributions and limitations, offering a clear, at-a-glance view of the current state of research in CubeSat cybersecurity. Furthermore, Fig.~\ref{fig:attacks_taxonomy} provides a structured visualization of the specific cyber threats and attack vectors discussed, mapped across the ground, link, and space segments.

\begin{table*}[!t]
\centering
\caption{Comparative Analysis of CubeSat Cybersecurity Efforts}
\label{tab:sec2-summary}
\resizebox{\textwidth}{!}{
\begin{tblr}{
  colspec = {X[l,2.2] X[l,3.0] X[l,3.4] X[l,3.6] X[l,4.6] X[l,5.2] X[l,5.2]},
  row{1} = {font=\bfseries},
  hlines,
  vlines,
}
Reference
& Category
& Platform / Environment
& Purpose
& Main Technique
& Key Contributions
& Limitations \\

\cite{attack_tree}
& Vulnerability Assessment and Attack Simulation
& Abstracted CubeSat architecture
& Assess CubeSat vulnerabilities such as denial of service, data tampering, and system disabling
& Attack tree analysis modeling attack vectors across ground, link, and space segments
& Systematic enumeration of attack pathways; structured methodology for evaluating CubeSat security posture
& Does not address practical implementation challenges or resource costs of countermeasures \\

\cite{hosted_payload}
& Vulnerability Assessment and Attack Simulation
& Emulated OPS-SAT payload environment
& Demonstrate a realistic low-resource cyberattack on a CubeSat camera payload
& Library hijacking of the camera executable, formalized with attack trees and SPARTA
& Demonstrated feasibility of a low-resource attack on an operational CubeSat
& Attack deployment method not explored, limiting reproducibility \\

\cite{HACK}
& Simulation Platforms for Cybersecurity Testing
& Conceptual Digital Twin of a CubeSat
& Simulate cyberattacks and analyze cascading multiphysics effects
& Model-based systems engineering with Digital Twin simulations
& Proposed a holistic framework for cyber-induced subsystem failure analysis
& Lacks practical guidance for Digital Twin development and implementation \\

\cite{linkstar}
& Simulation Platforms for Cybersecurity Testing
& LinkStar Cybersecurity Sandbox
& Enable hands-on cybersecurity testing across mission segments
& Physical testbed-based attack simulation using on-ground CubeSat hardware
& Provided a rare practical platform for cybersecurity experimentation
& Limited modeling of dynamic space environmental conditions \\

\cite{ids_survey}
& Intrusion Detection Systems
& N/A (survey study)
& Establish a taxonomy of IDS approaches for CubeSats
& Literature review and classification of IDS architectures and data sources
& Foundational taxonomy for CubeSat IDS research
& Analogies to non-space domains may not fully capture space-specific constraints \\

\cite{physics_based}
& Intrusion Detection Systems
& Conceptual CubeSat framework
& Integrate cyber and physical state for anomaly detection
& Correlation of cyber activity with physical state information
& Introduced an attack-agnostic, physics-informed IDS concept
& Conceptual proposal without implementation details \\

\cite{scenario_based}
& Intrusion Detection Systems
& Notional CubeSat model
& Generate realistic attack scenarios for IDS development
& Penetration testing-based scenario generation
& Produced detailed scenarios useful for IDS validation
& Limited applicability to diverse real CubeSat architectures \\

\cite{current_ids}
& Intrusion Detection Systems
& 3U CubeSat engineering model
& Design and validate an operational CubeSat IDS
& Two-stage ANN for CSP traffic over CAN bus, with the replay-attack classifier trained using the UAVCAN Attack Dataset
& Reported F1-scores of 99.59\% (DoS), 90.23\% (fuzzy injection), and 87.66\% (replay); represents the first practical implementation of a CubeSat IDS
& Trained on UAV-derived data, raising applicability concerns \\

\end{tblr}
}
\end{table*}

\begin{table*}[ht]
\centering
\caption{Comparative Analysis of Machine Learning Methods for Anomaly Detection in CubeSats (For studies using multiple algorithms, only the best-performing one is included in the table.)}
\label{tab:analysis}
\resizebox{\textwidth}{!}{
\begin{tblr}{
  colspec = {X[l,2.3] X[l,2.5] X[l,3] X[l,3] X[l,3] X[l,6] X[l,5.5]},
  row{1} = {font=\bfseries},
  hlines,
  vlines,
}
  Reference & Platform / Environment & Purpose & Algorithm & Data Sources & Findings & Limitations \\
  \cite{CaeBased} & LabSat (Simulated model) & ISHM tool for detecting, diagnosing, and explaining anomalies & Self-Organizing Maps (SOM) with Case-Based Reasoning (CBR) & Telemetry data from power generation, storage, and load consumption boards & High accuracy in anomaly detection and case retrieval; 94.4\% success in exact match retrieval with SOM-based indexing; effective in identifying known and novel anomalies. & Signature-based retrieval occasionally misclassified novel faults. The approach requires further validation on real system-generated data and optimization of case retrieval efficiency. \\
  \cite{solarPanel} & BIRDS-3 and BIRDS-4 & Detecting anomalies in CubeSat solar panel systems & Linear Discriminant Analysis (LDA) & Solar panel telemetry data: temperature, current, and voltage & LDA achieved the highest F1-Score (85\%); effective in anomaly detection with efficient resource utilization; challenges in handling imbalanced data. & ML models require adjustment for each mission's sensor calibration. The study focused only on solar panels; battery system faults were not analyzed. \\ 
  \cite{Horne2023859} & EduSat (flight-like model) & Improve on-board fault detection for CubeSats & Convolutional Neural Network (CNN) & Temperature telemetry data: sensors on maximum power point trackers, voltage converters, battery monitor & High accuracy in anomaly detection; effective handling of resource constraints; introduced holdoff window for improved reliability. & Performance varies across datasets due to noise; real-time inference introduces minor delays. \\ 
  \cite{mlDetects} & OPS-SAT & Detecting anomalies in satellite telemetry data & Random Forest & Magnetometer and Photodiode (PD) telemetry & High accuracy (98.4\%) in anomaly detection; addressed practical challenges of noisy and fragmented data; improved generalization with data augmentation. & Model relies on human-labeled training data, limiting adaptability to new anomalies. On-board deployment feasibility remains untested.\\ 
  \cite{TowardOn-boardDetection} & OPS-SAT & Detecting anomalies in satellite telemetry data & RNN-LSTM & Magnetometer and Photodiode (PD) telemetry & High accuracy (95.3\%) in anomaly detection; addressed practical challenges of noisy and fragmented data; effective manual annotation for training. & The model classifies anomalies but does not determine their onset, and may be sensitive to unseen telemetry patterns in real CubeSat missions. \\ 
  \cite{orbitAI} & OPS-SAT & On-board training for FDIR model using sensor data & AROW & Photodiode (PD) telemetry & Achieved 89\% balanced accuracy using the AROW algorithm; first instance of on-board training of AI models in space; addresses computational constraints and real-time data access. & Limited training input diversity restricted model generalization. High-dimensional models required more training iterations, and real-time constraints impacted performance.\\ 
\end{tblr}
}
\end{table*}

\section{Health and Telemetry Anomaly Detection in CubeSats}
\label{sec:noncyber}

CubeSats often experience mission failures in Low Earth Orbit (LEO) due to cost-effective components and simplified production processes. Anomaly detection systems are critical for identifying failures and mitigating risks such as collisions and space debris \cite{Horne2023859}. While traditional Out-of-limit (OOL) methods detect simple anomalies, they fall short for complex patterns. ML-based approaches provide a scalable, autonomous alternative, enabling more effective anomaly detection without constant expert intervention.

This section reviews anomaly detection methods applied to satellite health and telemetry data (such as power, thermal, and attitude control) rather than explicit cyber intrusions.
Although the studies in this section did not specifically address intrusion detection, the described algorithms and deployment methods are relevant for researchers focusing on CubeSat cybersecurity. Both physical and cyber IDSs share the need for efficient algorithms that can operate effectively within the resource constraints of CubeSats. Therefore, the methods mentioned in this section can be adapted to suit the detection of cyber anomalies by leveraging their efficiency and compatibility with CubeSat hardware.

The authors in \cite{CaeBased} presented the Anomaly Detection via Topological feature Maps (ADTM) system, tested on ``LabSat", a simulated model of the IT SPINS project. Designed as an Integrated System Health Management (ISHM) tool, the system combined Self-Organizing Maps (SOM) with Case-Based Reasoning (CBR), achieving a 94.4\% retrieval success rate. Focusing on power generation, the study in \cite{solarPanel} applied Linear Discriminant Analysis (LDA) to solar panel telemetry from the BIRDS-3 and BIRDS-4 CubeSats. The LDA algorithm demonstrated the best performance in this simulated environment with an F1-Score of 85\%. Similarly, \cite{Horne2023859} explored on-board anomaly detection using EduSat, an engineering model tested in a laboratory setting. This study found that Convolutional Neural Networks (CNN) offered superior performance for resource-constrained hardware, achieving F1-Scores up to 100\% across various test scenarios.

Significant advancements have been made using the OPS-SAT spacecraft. In \cite{mlDetects}, researchers validated anomaly detection algorithms using real magnetometer and photodiode telemetry from OPS-SAT. The Random Forest model achieved 98.4\% accuracy, proving effective on noisy, real-world signals. Taking this further, \cite{TowardOn-boardDetection} presented an end-to-end pipeline using Recurrent Neural Networks (RNN) with Long Short-Term Memory (LSTM) modules. Validated on the OPS-SAT development kit, this model achieved 95.3\% accuracy despite fragmented data. Finally, the OrbitAI application \cite{orbitAI} performed the first instance of on-board AI training. Using the Adaptive Regularization of Weights (AROW) algorithm for Fault Detection, Isolation, and Recovery (FDIR), the system achieved a balanced accuracy of 89\%, demonstrating the feasibility of online learning in space. A summary of the studies discussed above is shown in Table \ref{tab:analysis}.

The reviewed studies demonstrated advancements in health and telemetry anomaly detection for CubeSats, presenting a variety of ML approaches suitable for resource-constrained environments. Methods such as Self-Organizing Maps combined with Case-Based Reasoning (SOM-CBR) \cite{CaeBased}, LDA \cite{solarPanel}, CNNs \cite{Horne2023859}, and RNN-LSTMs \cite{TowardOn-boardDetection} demonstrated strong potential for autonomous anomaly detection. Additionally, the OrbitAI application represented an important step in on-board training using the AROW algorithm \cite{orbitAI}. These approaches addressed challenges such as noisy and fragmented telemetry data \cite{mlDetects}, but limitations remain, including dependency on human-labeled data, variability in performance across datasets, and insufficient diversity in training inputs for on-board models. The use of real telemetry data from OPS-SAT \cite{mlDetects, TowardOn-boardDetection, orbitAI} has advanced validation efforts, yet further testing in real operational scenarios is essential. As these techniques continue to evolve, their potential for adaptation to broader anomaly types, including cybersecurity, offers promising directions for future research.

\section{Existing Challenges in CubeSat Cybersecurity}
\label{sec:openresearch}
The importance of cybersecurity in space is starting to be recognized, as evidenced by studies on vulnerabilities and attack vectors in CubeSats \cite{attack_tree, hosted_payload, HACK}, and proposals and developments of environments such as Digital Twins and the LinkStar Cybersecurity Sandbox for testing cybersecurity vulnerabilities \cite{HACK, linkstar}. Furthermore, studies have outlined the necessary components of a CubeSat IDS \cite{ids_survey, physics_based, scenario_based}, and \cite{current_ids} has developed a practical implementation of IDSs for CubeSats. Despite these advancements, more research in several areas is required to implement resilient intrusion detection for CubeSats. These areas are discussed below.

\subsection{Evaluation and Testing of CubeSat IDS Solutions}
The only operational Intrusion Detection System developed specifically for CubeSats \cite{current_ids}, while promising, has not been thoroughly evaluated under realistic mission scenarios. In practice, a CubeSat may need to execute various complex tasks, such as Earth observation or inter-satellite communication, while simultaneously running IDS operations. for non-cybersecurity anomaly detection, practical experiments have demonstrated the feasibility of various ML approaches in CubeSats \cite{CaeBased, solarPanel, Horne2023859}. Additionally, significant advancements have been made with the use of real telemetry data for anomaly detection \cite{mlDetects, TowardOn-boardDetection, orbitAI}. However, these methods have not focused on detecting cybersecurity anomalies, and there remains a lack of empirical data showing how IDS deployments affect CubeSat performance during these mission-specific tasks. More extensive testing and data collection are required to determine the performance footprint of IDS solutions in realistic scenarios, ensuring that cybersecurity does not compromise mission objectives or exceed resource limitations.

\subsection{Resource-Efficient Intrusion Detection Mechanisms}
CubeSats face significant limitations in power, computational capacity, and communication bandwidth.

\begin{itemize}  
    \item{Power limitations:} The available power for CubeSat systems ranges from 2 to 8 watts, with the smallest and most cost-effective models at the lower end of this spectrum. This power capacity is achieved through the installation of solar panels, which are sized according to the CubeSat's compact dimensions of 10x10x10 cm \cite{power}.  
    \item{Computational limitations:} CubeSats rely on low-power microcontrollers, which typically provide only enough processing capability to operate the spacecraft, with little capacity for advanced sensor processing or autonomous functions. Even higher-performance processors face reliability challenges due to radiation susceptibility \cite{computation}.  
    \item{Bandwidth limitations:} Many CubeSats historically used amateur radio frequencies, particularly in the UHF band. Communication over UHF typically operates at just 9.6 Kbit/s, whereas higher-bandwidth solutions like the Wallops UHF Radar provide up to 3.0 Mbit/s. The crowded UHF band and reliance on low-data-rate transceivers further restrict performance. While some CubeSats use the S-band with a 24 Mbps downlink, modern missions require tens to hundreds of Mbps, potentially necessitating a shift to X-band for improved bandwidth and reduced interference \cite{bandwidth}.  
\end{itemize} 

To address these constraints, it is necessary to develop cybersecurity measures that are lightweight, require minimal memory and energy, and still provide effective threat detection. Balancing real-time anomaly detection with energy efficiency remains a challenge, and optimizing algorithms to operate within these limitations without compromising performance continues to be a key area of research \cite{ids_survey}.

\subsection{Autonomous Response Mechanisms for CubeSat Cybersecurity}
Unlike terrestrial systems, where the response to a detected threat can involve notifying a Security Information and Event Management (SIEM) center to coordinate a response, CubeSat IDS cannot always rely on real-time communication with a ground station due to limited line-of-sight \cite{los} contact. This highlights a need for research in developing autonomous security solutions that can handle threats without human intervention. This may involve entering a predefined safe state, rebooting critical systems, or attack-specific automated responses designed to minimize the impact of a threat until contact with a ground station can be re-established.

\subsection{Cybersecurity Frameworks for CubeSats}
Existing cybersecurity standards, often adapted from terrestrial systems, fail to address the unique challenges faced by space assets. Research must focus on creating specialized threat models that adequately consider the distinctive risks CubeSats face, including vulnerabilities across the supply chain, operational threats, and mission-specific challenges. The SPARTA matrix \cite{AerospaceSPARTA} categorizes attack techniques in space systems, representing a positive step forward. However, more work is needed to create a comprehensive framework that specifically secures CubeSats against emerging threats. CubeSats have a smaller attack surface due to resource constraints and a limited number of components. While this limits attackers' options, the low-power nature of CubeSats makes them vulnerable to energy-draining attacks, which could cause them to become inoperable \cite{attack_tree}.

\section{TinyML for CubeSat Cybersecurity}
\label{sec:tinyml}
The integration of Tiny Machine Learning (TinyML) into CubeSat cybersecurity presents an opportunity to address the resource constraints and dynamic challenges of these small satellites. This section explores TinyML techniques, their benefits, challenges, and potential applications of TinyML for CubeSat cybersecurity.

\subsection{What is TinyML?}
Tiny Machine Learning (TinyML) refers to the application of ML algorithms on small, resource-constrained devices like microcontrollers \cite{tinyML}. TinyML enables ML tasks to be performed directly on devices with extremely limited computational power, memory, and energy resources. These devices include microcontrollers and embedded systems, which are often constrained to milli- or microwatt power consumption \cite{tinyML_survey}.

\begin{figure}[htbp]
    \centering
    \includegraphics[width=\columnwidth]{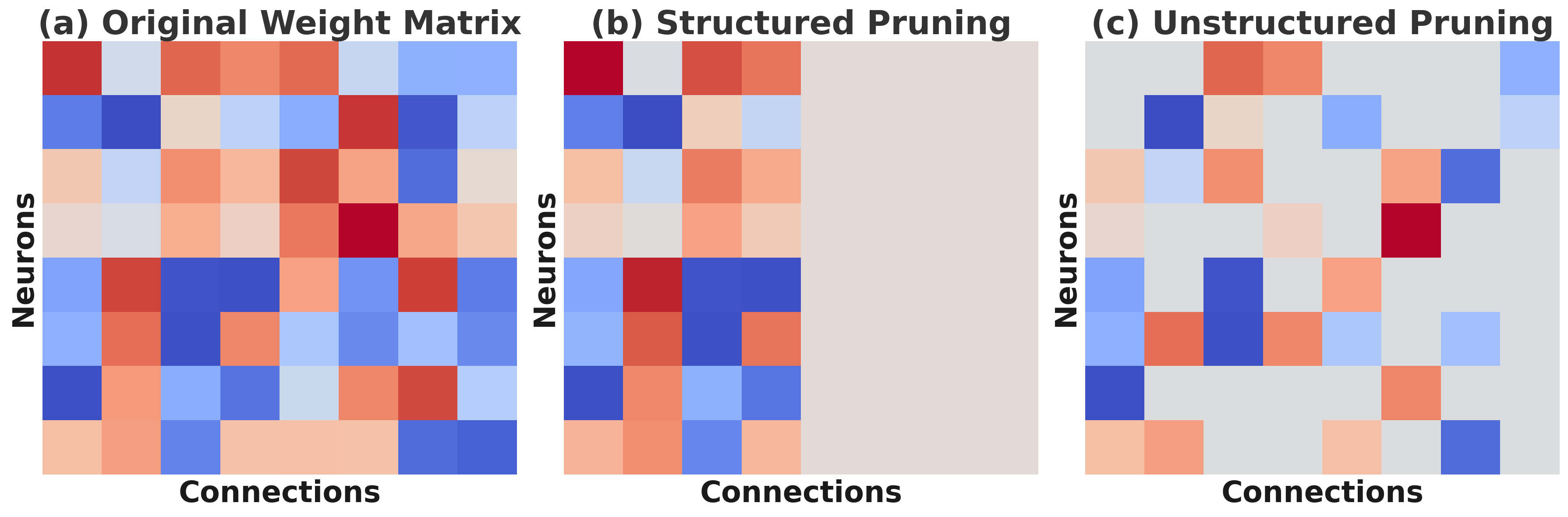}
    \caption{Example of pruning techniques applied to a neural network weight matrix. The left panel (a) shows the original weight matrix with all connections intact. The center panel (b) shows structured pruning, where entire columns representing neurons or filters are removed, producing regular sparsity. The right panel (c) shows unstructured pruning, where individual weights are zeroed out, producing irregular sparsity.}
    \label{fig:pruning_visualization}
\end{figure}

\begin{figure}[htbp]
    \centering
    \includegraphics[width=0.8\columnwidth]{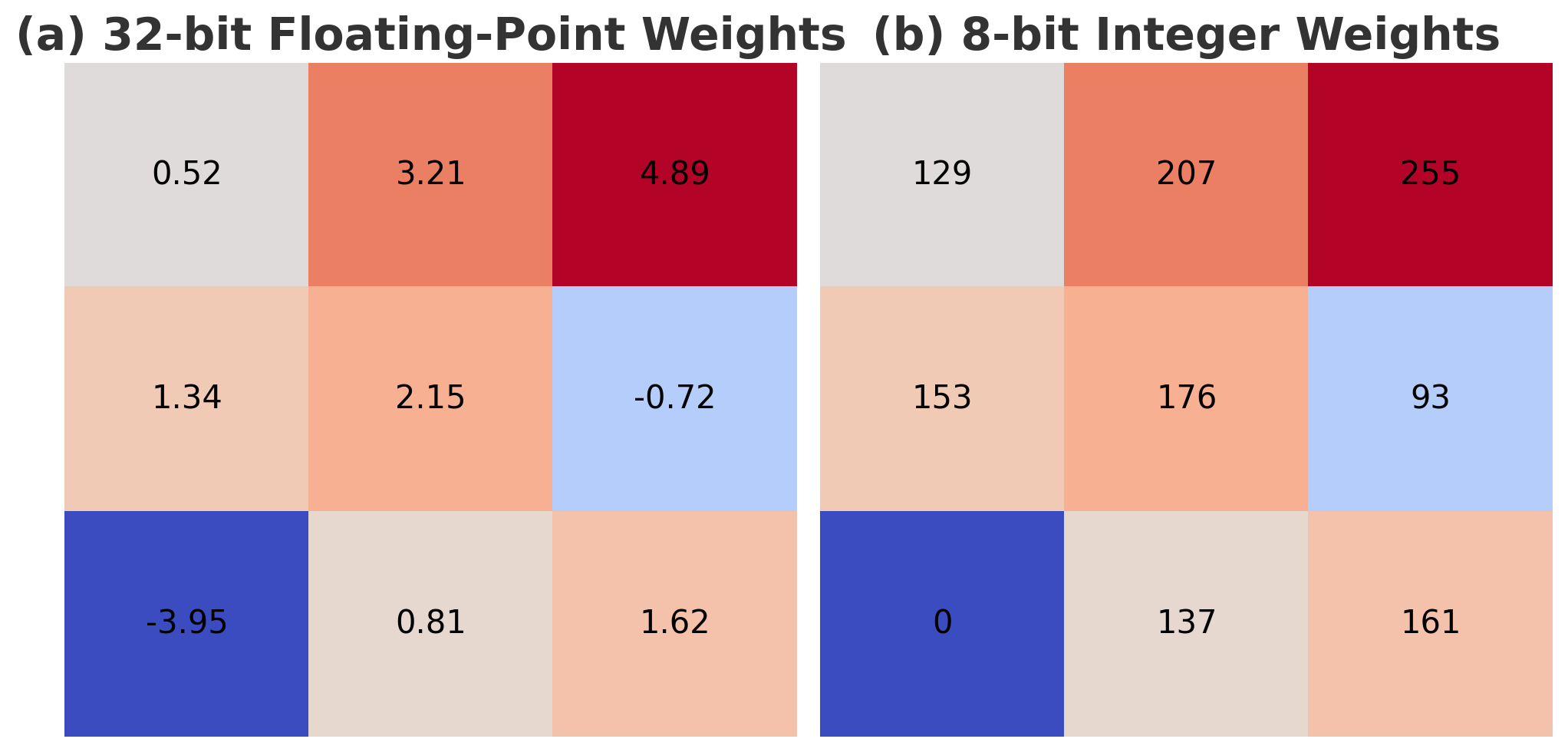}
    \caption{Example of the quantization process applied to neural network weights. The left panel (a) shows the original 32-bit floating-point weights with high precision. The right panel (b) shows the corresponding 8-bit integer weights produced using a uniform quantization method.}
    \label{fig:quant_visualization}
\end{figure}

\begin{figure}[!t]
    \centering
    \includegraphics[width=\columnwidth]{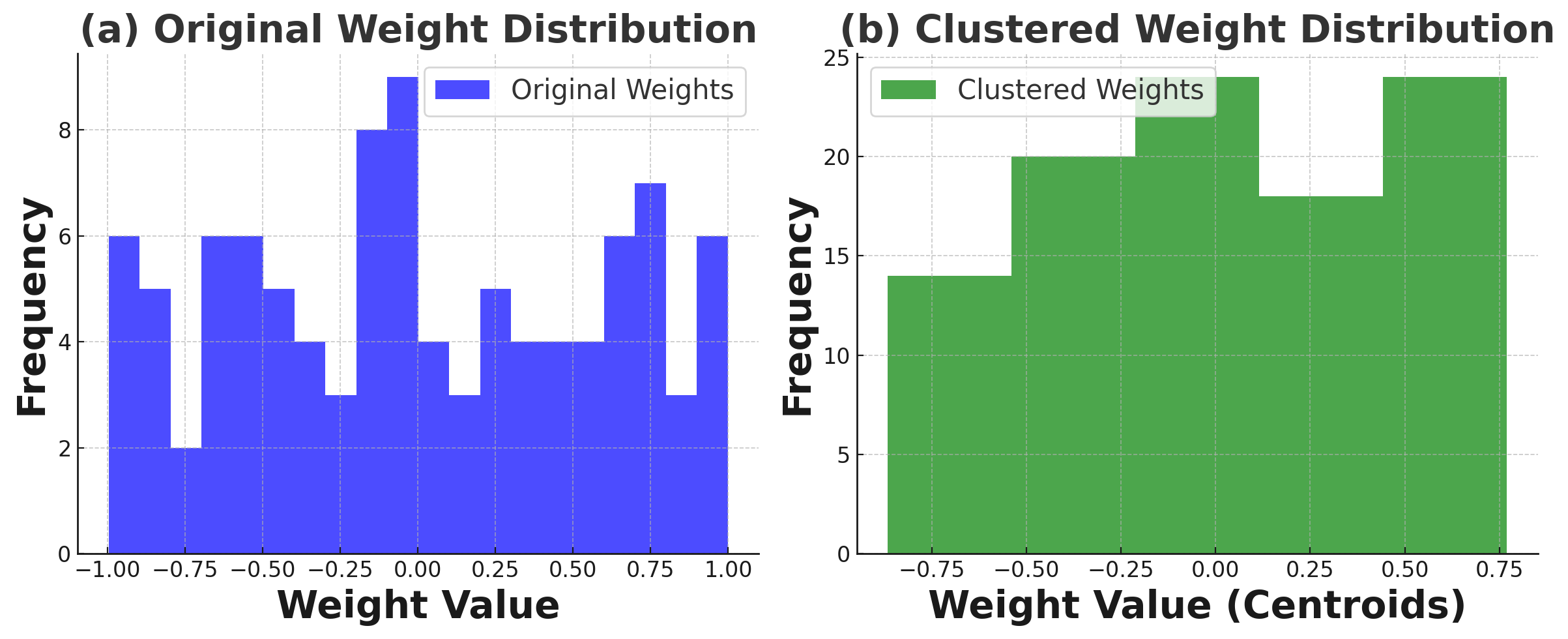}
    \caption{Example of weight clustering in a neural network. The left panel (a) shows the original weight distribution with a wide range of weight values. The right panel (b) shows the clustered weight distribution after grouping similar weights into a smaller set of centroid values using k-means clustering.}
    \label{fig:weight_clustering}
\end{figure}

At its core, TinyML uses techniques like pruning, quantization, and clustering to optimize ML models for resource-constrained devices. These techniques, designed to reduce the size, complexity, and computational demands of models, are described below:

\begin{itemize}
    \item Pruning: reduces the computational complexity of neural networks by removing unnecessary parameters. Structured pruning removes entire groups like channels or filters, improving inference speed, while unstructured pruning zeroes out individual weight connections, creating sparsity and reducing memory usage \cite{tinyML, structured}. An example of pruning is visualized in Fig. \ref{fig:pruning_visualization}.

    \item Quantization: a compression technique that reduces computational and memory demands by converting high-precision floating-point data into low-bit integer representations. For instance, a common practice is converting 32-bit floating-point weights and activations into 8-bit integers. By minimizing the bit-width of weights and activations, quantization achieves faster inference speeds and lower energy consumption \cite{quant}. Fig. \ref{fig:quant_visualization} illustrates how uniform quantization (with evenly spaced intervals) \cite{uniform} is applied to convert the original 32-bit floating-point weights into 8-bit integer values.

    \item Clustering: reduces the number of unique weight values in a neural network by grouping similar weights and replacing them with a smaller set of centroid values calculated from the original weights. This process, often performed using the k-means clustering algorithm, reduces memory usage and compresses the model, but compression should be validated on the target task since accuracy can decrease \cite{tinyMLDesc}. Fig. \ref{fig:weight_clustering} shows an example of weight clustering in a neural network, where an original distribution of unique weight values is transformed into a smaller set of centroid values using k-means.
\end{itemize}

These optimizations enable TinyML to operate effectively on low-power devices like CubeSats.

\subsection{Advanced TinyML Technologies}

While basic TinyML techniques provide foundational improvements, advanced methods are required to enhance adaptability, efficiency, and security in resource-constrained environments. This section explores key advancements, including optimized model architectures, transfer and continual learning, federated learning, dimensionality reduction, and quantization-aware training. These techniques, summarized in Table \ref{tab:advancedtinyMLtech}, improve model efficiency, adaptability, and execution, allowing CubeSats to process data reliably despite strict power and memory limitations.

\begin{table*}[htbp]
\centering
\caption{Summary of Advanced TinyML Techniques}
\label{tab:advancedtinyMLtech}
\resizebox{\textwidth}{!}{%
\begin{tblr}{
  colspec = {X[1.7,c] X[2.5,l] X[4,l] X[3,l]},
  row{1} = {font=\bfseries},
  hlines,
  vlines,
}
\textbf{Technique} & \textbf{Description} & \textbf{TinyML Adaptation} & \textbf{CubeSat Relevance} \\

\textbf{Efficient Model Optimization, NAS, and CASH} &
Optimizes ML models for constrained devices by reducing computation and memory while maintaining accuracy. NAS automates architecture search under resource constraints, and CASH jointly selects a model family and its hyperparameters within a single search space \cite{lin2021efficient,thornton2013auto}. &
Microcontroller-aware NAS like TinyNAS can explicitly account for constraints such as memory and latency, and can be paired with memory-efficient inference runtimes \cite{lin2021efficient}. CASH can also be used as a TinyML tuning step when its objectives are defined to reflect deployment needs, so the selected model family and hyperparameters balance detection quality with inference cost and footprint. &
Supports on-board deployment by balancing detection performance, inference responsiveness, and model footprint under strict CubeSat resource limits. \\

\textbf{Continual and Transfer Learning (TinyTL, TinyOL)} & Adapts models to new data post-deployment with minimal retraining, reducing training resource consumption \cite{10284551}. & TinyTL fine-tunes certain parameters (e.g., biases) while freezing most weights to reduce memory and computation overhead \cite{tinyMLDesc}. TinyOL continuously updates models in real-time using small incremental updates \cite{tinyol}. & Enables CubeSats to refine on-board ML models and adapt to dynamic conditions without requiring frequent ground updates. \\

\textbf{Federated Learning (FL)} & Enables decentralized training across multiple devices without sharing raw data, preserving privacy \cite{kopparapu2022tinyfedtl}. & TinyFedTL enables federated learning on microcontrollers with less than 1MB of memory by optimizing communication and model update strategies \cite{kopparapu2022tinyfedtl}. & Reduces communication costs and enhances privacy by allowing CubeSats to collaboratively train models while keeping on-board data secure. \\

\textbf{Automated Feature Selection (AutoFS)} &
Automatically selects a subset of informative features to remove irrelevant or redundant variables and reduce dimensionality \cite{yang2022iot}. &
AutoFS can be used as a TinyML feature-pruning step when its objectives reflect deployment needs, selecting a compact feature subset that preserves detection capability while reducing the number of features that must be computed on board at inference time. &
Reduces the number of real-time feature computations required on board and lowers the input footprint, improving feasibility under CubeSat compute, memory, and storage limits \cite{saha2022machine}. \\

\textbf{Quantization-Aware Training (QAT)} & Simulates quantization effects during training to maintain accuracy after reducing precision \cite{10284551}. & QAT ensures models remain robust at lower bit precision (e.g., 8-bit) by incorporating quantization constraints into training, minimizing accuracy loss \cite{10284551}. & Helps CubeSats maintain model accuracy despite precision fluctuations. \\
\end{tblr}
}
\end{table*}

\subsubsection{Efficient Model Optimization with Neural Architecture Search (NAS) and Combined Algorithm Selection and Hyperparameter (CASH)}
Deploying ML models on resource-limited devices like CubeSats requires optimization to balance performance, memory usage, and energy efficiency. This motivates TinyML model optimization techniques that reduce complexity while maintaining high accuracy.

One approach is lightweight model architectures, which are specially designed to use fewer computational resources than conventional deep learning models. Many widely used architectures were designed for server or mobile-class hardware and can remain too large for microcontrollers, even after quantization. To address this, TinyML workflows rely on lightweight architectures and compression strategies such as depthwise separable convolutions, pruning, and sparsity to reduce computation and memory while preserving accuracy \cite{lin2021efficient}.

Beyond manually designing these optimized models, Neural Architecture Search (NAS) automates the process of finding the best possible architecture for a given hardware constraint. Microcontroller-aware NAS methods such as TinyNAS \cite{lin2021efficient} explicitly account for constraints such as memory and latency, and can be paired with memory-efficient inference runtimes to reduce peak memory usage during execution.

\begin{figure}[ht]
  \centering
  \includegraphics[width=0.9\linewidth]{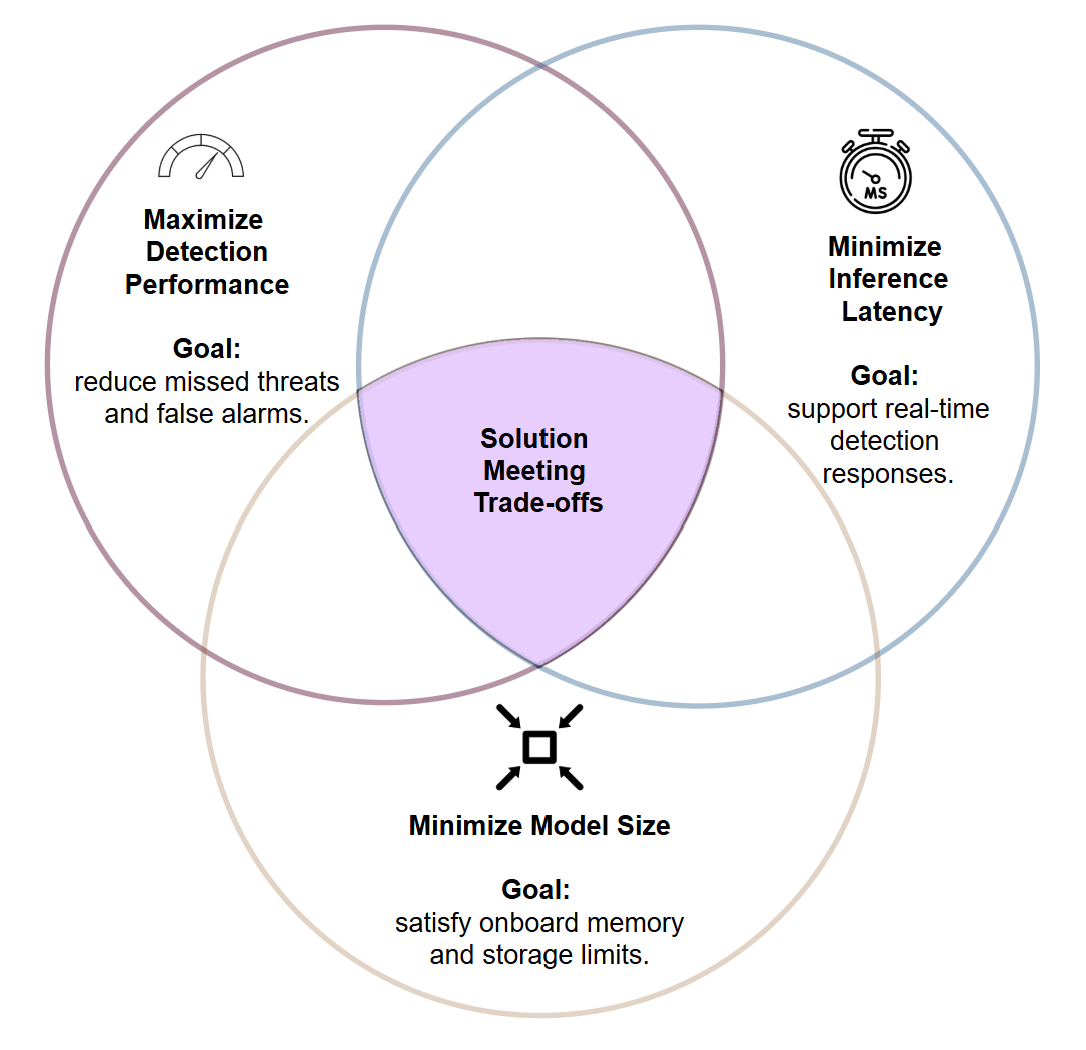}
  \caption{Illustrative CASH trade-off for TinyML deployment, balancing detection performance, inference latency, and model footprint to obtain a configuration that meets on-board constraints.}
  \label{fig:cash_tradeoffs}
\end{figure}

In addition to architecture-level optimization, Combined Algorithm Selection and Hyperparameter Optimization (CASH) jointly searches over the learning algorithm family and its hyperparameters as a single hierarchical optimization problem, where the algorithm identity is treated as a top-level hyperparameter and only activates the corresponding conditional hyperparameters when selected \cite{thornton2013auto}. With appropriate objectives, this joint search can be treated as a TinyML optimization technique because it can explicitly target deployability constraints alongside detection quality. As an illustrative example, CASH can be expressed as a joint search over an algorithm choice $a \in \mathcal{A}$ and its conditional hyperparameters $\lambda \in \Lambda_{a}$:
\begin{equation}
\max_{a \in \mathcal{A},\, \lambda \in \Lambda_{a}} \;\; \big( J_{\text{det}}(a,\lambda),\; -J_{\text{lat}}(a,\lambda),\; -J_{\text{size}}(a,\lambda) \big),
\end{equation}
where $J_{\text{det}}$ captures detection quality on validation data, $J_{\text{lat}}$ captures inference-time cost, and $J_{\text{size}}$ captures a footprint proxy such as serialized model size. As shown in Fig.~\ref{fig:cash_tradeoffs} such formulation can return configurations that meet deployment trade-offs rather than optimizing a single metric in isolation.

These techniques enable CubeSats to execute complex models within strict power and memory limits. With optimized architectures, NAS-based constraint-aware design, and CASH-based tuning, CubeSats can support on-board inference while keeping latency and model footprint within practical limits.

\subsubsection{Transfer and Continual Learning for Dynamic Adaptation}
TinyML enables models to adapt over time by learning from new data, even after deployment. This is particularly important for resource-constrained devices like CubeSats, which must operate in changing environments where pre-trained models may become outdated. To achieve continual adaptation, TinyML employs Tiny Transfer Learning (TinyTL) and Tiny Online Learning (TinyOL), designed to update models efficiently while minimizing memory and computation overhead \cite{10284551}.

TinyTL is a technique that optimizes transfer learning for resource-constrained devices by reducing memory usage during model fine-tuning. Instead of updating all model parameters, TinyTL freezes the weights and only fine-tunes biases, allowing intermediate activations to be discarded during backpropagation. This reduces peak memory usage and computation overhead, making it feasible for microcontrollers to adapt pre-trained models to new tasks without requiring extensive retraining \cite{tinyMLDesc}.

TinyOL enables real-time model adaptation by updating parameters gradually as new data arrives. Unlike traditional learning methods that require large batches of data and extensive retraining, TinyOL continuously fine-tunes a model with small updates, allowing it to learn dynamically without requiring high computational power \cite{tinyol}.

Together, TinyTL and TinyOL make it possible for TinyML models to adapt while remaining within the strict memory and energy constraints of CubeSats. They enable mission-long learning for space-based systems, allowing them to update their models without relying on frequent ground station interventions.

\subsubsection{Federated Learning and Privacy-Preserving Model Training}
Federated Learning (FL) is a ML approach that allows multiple devices to collaboratively train a shared model without exchanging raw data. Instead of centralizing data on a single server, each device performs local training and only shares model updates with a central aggregator, preserving data privacy while reducing communication overhead.

In TinyML, FL enables on-device learning for resource-constrained systems like microcontrollers by optimizing communication and memory usage. Traditional FL methods, such as Federated Averaging (FedAvg), have been widely used in mobile and cloud-based applications but struggle with the severe hardware limitations of TinyML devices. To address this, TinyFedTL, a specialized FL framework, has been developed to enable federated transfer learning on devices with less than 1MB of memory. TinyFedTL optimizes training by limiting memory usage, reducing communication frequency, and implementing efficient model update strategies. Unlike previous FL approaches, which rely on large-scale embedded systems, TinyFedTL successfully applies FL to highly constrained devices, demonstrating its feasibility for real-world IoT applications \cite{kopparapu2022tinyfedtl}.

For CubeSats, FL offers a promising solution to enable collaborative learning across a constellation of satellites while preserving on-board data privacy and minimizing communication costs. By training models locally on each CubeSat and only exchanging updates, FL reduces reliance on continuous ground station connectivity, making autonomous adaptation to space conditions more feasible. However, implementing FL in CubeSats introduces challenges such as intermittent connectivity, synchronization issues, and power constraints, which require further research to optimize FL for space-based systems.

\subsubsection{Automated Feature Selection for Efficient Model Execution}
Efficient TinyML deployment can benefit from reducing the input feature space so models remain feasible on microcontroller-class hardware with tight compute, memory, and energy budgets \cite{saha2022machine}. In this context, feature selection is a pre-processing step that chooses a subset of the original variables to remove irrelevant information and reduce data dimensionality \cite{yang2022iot}. For on-board CubeSat IDS use, this approach supports deployability because fewer selected features typically mean fewer feature computations at inference time, so the flight software extracts or computes fewer quantities in real time before passing them to the model.

AutoFS extends this idea by automating the search for a suitable subset rather than relying on manual trial-and-error. Common strategies can be grouped into: (i) filter methods, which score features using criteria that are independent of the learning algorithm; (ii) wrapper methods, which generate candidate subsets and evaluate them using a learning algorithm; and (iii) embedded methods, which integrate selection into training via the model's objective and implicitly choose features \cite{yang2022iot}. For example, one way to express AutoFS at a high level is as an optimization over subsets $S \subseteq \mathcal{F}$, where $\mathcal{F}$ is the full feature index set:
\begin{equation}
\max_{S \subseteq \mathcal{F}} \;\; \big( J_{\text{det}}(S),\; -J_{\text{lat}}(S),\; -J_{\text{foot}}(S) \big),
\end{equation}
where $J_{\text{det}}(S)$ captures detection quality on validation data, $J_{\text{lat}}(S)$ captures inference time, and $J_{\text{foot}}(S)$ captures a footprint proxy such as feature count or memory use. Framing AutoFS this way makes the trade-offs explicit: it can prioritize strong detection while simultaneously constraining latency and on-board resource usage for real-time, resource-limited deployment.

\subsubsection{Quantization-Aware Training for High-Accuracy Inference}
Quantization reduces model size and computational demands by encoding neural network weights and activations with lower precision (e.g., 8-bit instead of 32-bit floating point). However, naïve post-training quantization can lead to significant accuracy drops due to rounding errors and loss of information. Quantization-Aware Training (QAT) addresses this by simulating quantization effects during training, allowing the model to learn robust representations despite reduced precision. QAT works by incorporating quantization constraints directly into the training process, making sure that weight updates account for the eventual reduced precision of the deployed model. This minimizes performance degradation and allows TinyML models to maintain higher accuracy while benefiting from smaller memory footprints and faster inference speeds \cite{10284551}.
In the context of CubeSats, by training models to be inherently robust to precision reduction, QAT can improve resilience throughout the mission lifecycle.

\subsection{Benefits and Challenges of TinyML in CubeSat Settings}

TinyML offers several advantages that align well with the operational constraints of CubeSats. One major advantage is its energy-saving capability \cite{dehrouyeh2024tinyml}. CubeSats operate on restricted power supplies, often depending on solar panels with limited output. TinyML's design allows critical computations to occur directly on low-energy hardware like microcontrollers, reducing the energy drain caused by repeatedly sending data to ground stations or other satellites. This efficient use of power extends mission durations by preserving battery life. Another benefit is latency reduction. By processing data locally, TinyML enables CubeSats to analyze and act on information instantly without waiting for Earth-based systems for decision-making. For cybersecurity, this enables faster identification of anomalies and immediate responses to potential breaches or attacks. Additionally, privacy and security are enhanced since sensitive data remains on-board, reducing the risk of eavesdropping during transmission \cite{tinyML_survey}. By processing telemetry and operational data on-board, TinyML provides an extra layer of defense against interception. Reliability and availability are also key benefits. TinyML enables CubeSats to operate autonomously without continuous reliance on communication with ground stations, which can be disrupted by orbital conditions or limited communication windows. This autonomy allows CubeSats to maintain consistent cybersecurity monitoring, even in isolated environments.

\begin{table*}[htbp]
\centering
\caption{Benefits of TinyML for CubeSat Deployment}
\label{tab:tinyml_benefits}
\resizebox{\textwidth}{!}{%
\begin{tblr}{
  colspec = {X[1,c] X[5,l]},
  row{1} = {font=\bfseries},
  hlines,
  vlines,
}
\textbf{Benefit} & \textbf{Description and Citations} \\

\textbf{Energy Efficiency} & TinyML enables critical computations directly on low-power hardware like microcontrollers, reducing energy consumption. By limiting the need for frequent data transmissions to ground stations, TinyML conserves battery life and extends mission durations \cite{dehrouyeh2024tinyml}. \\

\textbf{Latency Reduction} & Localized data processing allows CubeSats to analyze and act on information instantly, minimizing delays. This is particularly beneficial for cybersecurity, enabling faster intrusion detection and immediate responses to threats. \\

\textbf{Privacy and Security} & TinyML keeps sensitive data on-board, reducing the risk of interception during transmission. Since it processes data locally, TinyML enhances data privacy and strengthens security against eavesdropping \cite{tinyML_survey}. \\

\textbf{Autonomy and Reliability} & TinyML allows CubeSats to operate independently without relying on constant communication with ground stations. This ensures consistent performance, even when communication windows are limited or disrupted by orbital conditions. \\
\end{tblr}
}
\end{table*}

\begin{table*}[htbp]
\centering
\caption{Challenges of TinyML for CubeSat Deployment}
\label{tab:tinyml_challenges}
\resizebox{\textwidth}{!}{%
\begin{tblr}{
  colspec = {X[1,c] X[5,l]},
  row{1} = {font=\bfseries},
  hlines,
  vlines,
}
\textbf{Challenge} & \textbf{Description} \\

\textbf{Real-Time Inflexibility} & TinyML models are typically pre-trained offline and deployed as static frameworks on edge devices \cite{dehrouyeh2024tinyml}. In dynamic CubeSat environments, data distribution changes over time (concept drift), leading to ineffective threat detection for novel attack patterns. Implementing adaptive techniques like TinyOL may mitigate this issue but requires additional computational resources \cite{conceptdrift, tinyol}. \\

\textbf{Model Robustness} & Harsh space conditions (e.g., radiation, extreme temperatures, electromagnetic interference) can compromise model performance on CubeSats. These perturbations reduce accuracy in functions like intrusion detection \cite{dehrouyeh2024tinyml}. Training models with noise perturbations during development can improve resilience \cite{ning2021improving}. However, accurately simulating all conditions for CubeSats may not be feasible. \\

\textbf{Accuracy Drop} & Compression and optimization techniques used in TinyML (e.g., quantization and pruning) can lead to slight accuracy reductions, which may impact critical operations \cite{tinyML_survey}. Extensive testing in high-fidelity simulation environments can ensure accuracy remains within acceptable limits under space-like conditions. \\

\textbf{Computational Constraints} & CubeSats operate with severe memory, power, and processing limitations. Implementing adaptive techniques like TinyOL or other complex algorithms often requires balancing computational cost with real-time performance \cite{tinyol}. \\
\end{tblr}
}
\end{table*}

Despite its benefits, TinyML also presents various challenges that must be addressed for effective CubeSat deployment. One major challenge is real-time inflexibility. TinyML models are typically trained offline on high-performance systems using pre-existing datasets and then deployed as static frameworks on edge devices \cite{dehrouyeh2024tinyml}. While effective in controlled settings, CubeSats operate in highly variable space environments where cyber threats can emerge or evolve unexpectedly. Static models lack the ability to update dynamically, leaving systems vulnerable to novel attack patterns or shifting operational conditions. This rigidity limits their effectiveness in providing timely threat detection and mitigation. The dynamic nature of CubeSat environments means the data distribution may change over time, a problem known as concept drift \cite{conceptdrift}. Static TinyML models struggle to adapt to such shift, leading to their inability to detect novel threats. To address this limitation, mentioned approaches such as TinyOL are being explored \cite{tinyol}. Implementing techniques like TinyOL on CubeSats would require consideration of additional computational resources to ensure efficient updates without compromising performance.

Another challenge is model robustness which describes a model's ability to preserve its performance when exposed to noise or disturbances \cite{dehrouyeh2024tinyml}.
CubeSats are exposed to harsh conditions such as radiation, extreme temperatures, and varying electromagnetic interference, all of which can compromise the performance of on-device TinyML models. These environmental perturbations can lead to inaccuracies in anomaly detection or cybersecurity functions. One way to address this is by training models with noise perturbations during development to improve their resilience against such disturbances \cite{ning2021improving}. For CubeSats, this approach involves incorporating representative perturbations during training, recognizing that simulating every possible space condition may not be feasible. Accuracy degradation is also a concern. TinyML models undergo compression and optimization to fit the limited computational capacity of edge devices, leading to a slight reduction in accuracy \cite{tinyML_survey}. To address this, extensive testing in a high-fidelity digital twin or simulated environment that mimics the harsh conditions and operational scenarios of CubeSats can help determine whether the accuracy drop remains within acceptable limits.

Understanding and addressing these benefits and challenges (summarized in Tables \ref{tab:tinyml_benefits} and \ref{tab:tinyml_challenges} respectively) is essential for establishing a dependable and resilient framework to protect CubeSats in the extreme and dynamic space environment.

\subsection{How Can TinyML Be Used to Enhance CubeSat Cybersecurity?}

TinyML offers practical applications for improving CubeSat cybersecurity by enabling on-board analysis of telemetry data (e.g., attitude control, thermal readings) and cyber state information (e.g., CPU usage, system logs). By processing this data locally, TinyML allows CubeSats to detect potential issues and respond autonomously, even when communication with ground stations is limited.

One potential application of TinyML is in analyzing telemetry data to identify patterns that may indicate potential problems. For example, unexpected changes in power consumption or temperature levels could indicate hardware malfunctions or external disruptions. TinyML could help detect such issues in real time, supporting timely responses. Another possibility involves monitoring communication traffic to identify irregularities or suspicious activity. TinyML could analyze patterns in data exchanges, flagging unexpected transmission spikes or unauthorized communication attempts. This capability would enhance CubeSat resilience by addressing potential vulnerabilities in its communication systems. By integrating both telemetry and communication data, TinyML could also be used to develop a host-based intrusion detection system tailored for CubeSats. This IDS would offer a more comprehensive understanding of the CubeSat's operational environment, merging insights from its physical and cyber states. Once a potential threat is detected, the CubeSat could take a threat-specific action or enter a general predefined safe state to minimize risks. It could then notify the ground station during the next communication window.

\begin{figure}[htbp]
\centering
\includegraphics[width=\linewidth]{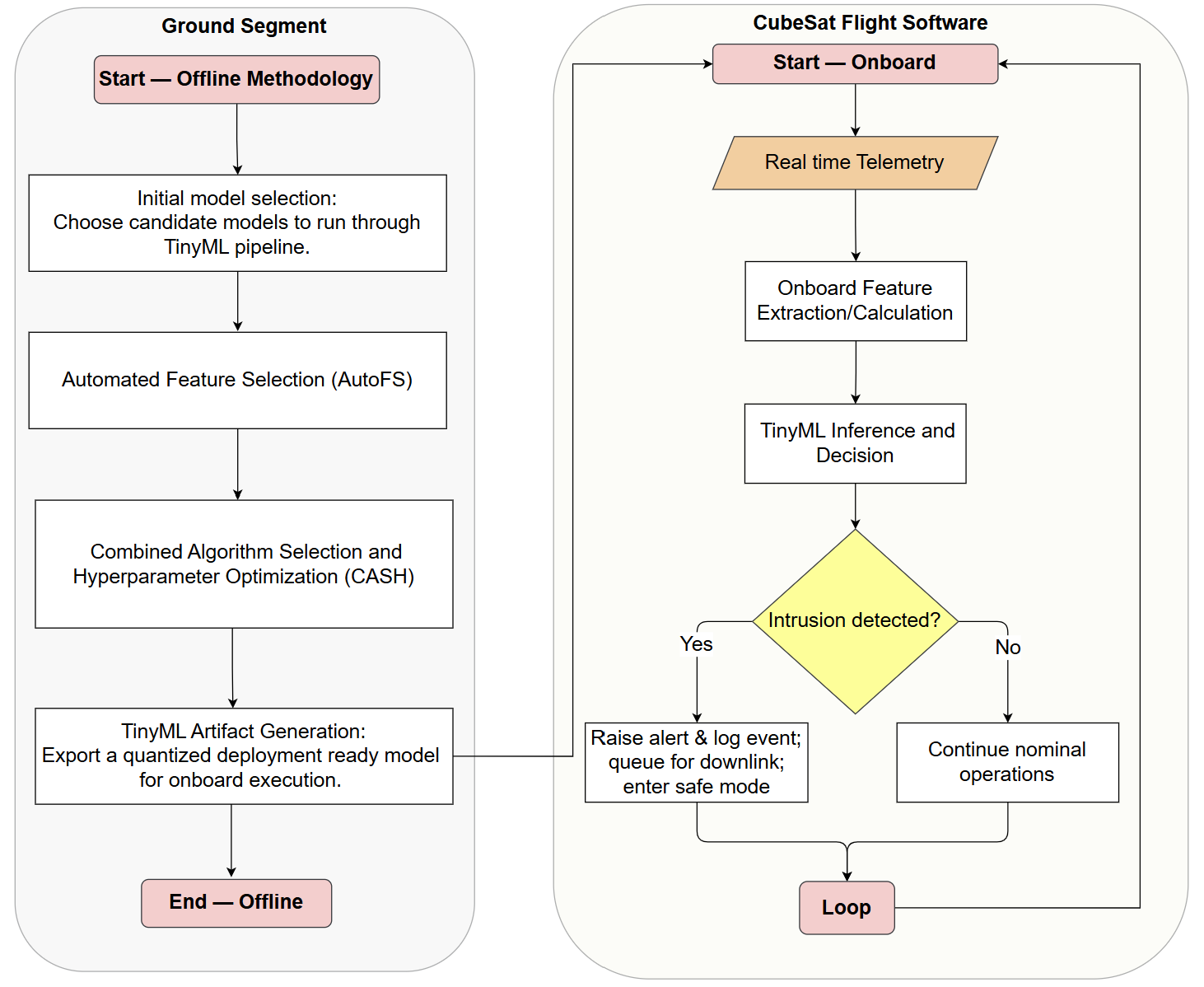}
\caption{Example TinyML-enabled Host-based IDS Workflow in a CubeSat Setting.} 
\label{fig:tinyml_hids_workflow}
\end{figure}

Fig.~\ref{fig:tinyml_hids_workflow} presents an example TinyML-enabled host-based IDS workflow that separates the offline ground-segment methodology from the on-board execution loop. Offline, the pipeline first selects candidate model families to run through the TinyML pipeline. AutoFS is then applied on a per-model basis to rank and retain a compact feature set that balances efficiency constraints with strong detection for that specific model family, preserving the features it finds most informative. CASH then jointly tunes the selected model type and its hyperparameters before exporting a quantized, deployment-ready TinyML artifact. On-board, the CubeSat processes real-time telemetry, performs feature extraction and calculation, and runs TinyML inference to decide whether an intrusion is detected. If an intrusion is detected, the system raises an alert and logs the event, then queues it for downlink and may enter a predefined safe mode. If no intrusion is detected, the CubeSat continues nominal operations and repeats the loop.

Developing these capabilities requires training TinyML models with representative data that includes both nominal operational conditions and scenarios involving potential threats. Techniques such as on-device learning could also be explored to allow these models to adapt over time, ensuring effectiveness in dynamic and evolving space environments.

\section{Other Future Directions for CubeSat Cybersecurity}
\label{sec:future}
The growing complexity of CubeSat missions, combined with the rise in cybersecurity threats, calls for innovative solutions to ensure effective anomaly detection and system protection. While TinyML has emerged as a promising approach for addressing CubeSat resource constraints and enabling efficient intrusion detection, several other areas remain unexplored. This section explores various promising research directions that can complement TinyML advancements and further improve CubeSat security by addressing the unique limitations of these small satellites.

\subsection{Integrating Cybersecurity with Health Monitoring for CubeSats}
Research on health monitoring systems for CubeSats is relatively mature compared to intrusion detection methods. Several studies have focused on developing anomaly detection techniques for monitoring CubeSat systems \cite{CaeBased, solarPanel, Horne2023859, mlDetects,TowardOn-boardDetection}, with an emphasis on maintaining mission reliability and minimizing the likelihood of failure. Given the limited resources available on CubeSats, developing standalone IDSs may be inefficient. A more practical solution is to integrate cybersecurity mechanisms into the existing health monitoring systems. By sharing computational resources, the system could manage both physical and cyber threats without significantly increasing the resource burden on the CubeSat. In addition, combining both the physical and cyber states of the satellite would provide a comprehensive understanding of its condition, allowing the detection of cyber anomalies that might not be apparent when considering the cyber state alone \cite{physics_based}.

\subsection{Collaboration Between Cybersecurity Researchers and Space Domain Experts}
The development of practical cybersecurity solutions for CubeSats has lagged behind the progress seen in health monitoring systems. One potential reason for this gap is that cybersecurity researchers often lack the specialized knowledge required to understand CubeSat systems comprehensively. This specialized knowledge is crucial for securing space assets, as understanding their unique design and functionality is essential to mitigate the associated cybersecurity risks effectively.

The structure of many space system organizations can exacerbate this issue. Security teams often lack the expertise to distinguish between traditional IT infrastructure and the specialized nature of satellite or ground control systems. This lack of differentiation means that cybersecurity efforts for space-specific assets can be deprioritized. Moreover, these teams frequently face broad responsibilities across both IT and space domains, which overstretches their resources and increases the vulnerability of space assets \cite{vacuum}.

To address this challenge, collaboration between cybersecurity researchers, ethical hackers, and domain experts with deep knowledge of CubeSat systems is necessary. Such collaboration could help develop solutions that consider both cybersecurity concerns and space-specific operational requirements. Another potential outcome of such collaboration could be the creation of realistic datasets that can be used to develop cybersecurity solutions for CubeSats.

Simulation platforms such as the LinkStar Cybersecurity Sandbox \cite{linkstar} provide one approach for creating these datasets. This platform allows researchers to simulate potential cyberattacks across various segments of a satellite mission and gather data specific to cyber intrusion attempts. However, while these simulations provide controlled and reproducible environments for collecting cybersecurity data, they may not fully capture the complexities and environmental conditions experienced by CubeSats in space.

Another method for collecting data is through real-world space missions, which offer valuable opportunities to obtain data directly from operational CubeSats. For instance, a telemetry dataset from the OPS-SAT mission known as OPSSAT-AD \cite{opsad} provides annotated data for anomaly detection, including features derived from real telemetry signals. While this dataset primarily focuses on operational anomalies, a future mission could expand its scope to include cybersecurity-specific scenarios. This would require collaboration between domain experts and cybersecurity researchers to simulate and document cyberattacks on CubeSats during their operational lifecycle. However, the financial and logistical challenges associated with launching and operating CubeSats for such specialized purposes must be carefully considered. Additionally, the time required to design, execute, and analyze real space missions could limit the pace of cybersecurity research.

Additionally, existing cybersecurity datasets like the CICIDS2017 dataset\cite{cicds}, which contains network intrusion data, can be adapted for CubeSat environments. Expert knowledge can be used to map the dataset to represent space communication protocols such as the Space Packet Protocol (CCSDS 133.0-B-2) \cite{CCSDS1330B2}. This adaptation would involve translating terrestrial network behaviors and potential attack patterns to scenarios relevant to satellite operations. While this approach provides a cost-effective and scalable option, it may not fully replicate the conditions and complexities of space-based communication systems.

\section{Conclusion}
CubeSats have revolutionized space accessibility but remain highly vulnerable to cybersecurity threats due to their reliance on COTS components and open-source software. As CubeSats take on expanded roles, especially in national security, addressing these vulnerabilities is critical. Current cybersecurity efforts involve tools like attack tree analysis, Digital Twins, and the LinkStar Cybersecurity Sandbox, which simulate potential threats and evaluate vulnerabilities. The development of a CubeSat-specific intrusion detection system represents significant progress, though further testing under realistic mission conditions remains necessary.

Non-cyber anomaly detection methods, such as health monitoring and fault detection, have proven effective using algorithms like Self-Organizing Maps (SOM), Linear Discriminant Analysis (LDA), and Recurrent Neural Networks (RNNs). While these techniques primarily focus on physical health, they offer valuable insights for adapting ML strategies to intrusion detection in resource-constrained environments. Achieving resilient intrusion detection for CubeSats requires addressing open questions, including evaluating IDS in real missions, designing resource-efficient mechanisms, and developing autonomous systems that operate without ground station reliance. TinyML offers a promising solution to these challenges by enabling real-time, low-power intrusion detection directly on CubeSats. By processing data locally, TinyML reduces latency and enhances cybersecurity. Challenges such as static model deployment and environmental robustness persist, but techniques such as TinyOL and robust training methods show significant potential for CubeSat applications.

Future research should integrate cybersecurity with health monitoring systems to address both physical and cyber threats, while optimizing limited resources. Collaboration between cybersecurity and space experts is essential to creating tailored solutions for CubeSats.

Building on the design patterns and constraints synthesized in this survey, our next step is to implement and evaluate an end-to-end TinyML-based IDS pipeline in a CubeSat testbed, reporting quantitative trade-offs among detection performance, inference latency, and model footprint. Strengthening CubeSat cybersecurity will enhance resilience, safeguard investments, and promote the sustainability of space operations as these satellites take on increasingly critical roles in space missions.

\bibliographystyle{ieeetr}
\bibliography{sample-base}

\end{document}